**Specific propagation directions for media of rhombic symmetry**

Artur Duda, Institute of Theoretical Physics, University of Wroclaw, pl. Maxa Borna 9,

**Abstract**

We reanalyze the problem of existence of longitudinal normals inside symmetry planes of piezoelectric crystals belonging to the symmetry class mm2. The equations determining components of longitudinal normals situated outside symmetry planes for media of this symmetry are discussed. It is proven that non-piezoelectric media of rhombic symmetry could have 4 or 8 distinct acoustic axes. Examples of non-piezoelectric elastic media of monoclinic symmetry without acoustic axes are given. The method of determination of components of acoustic axes for piezoelectric media of arbitrary symmetry is presented. With the help of this method we discuss the problem of acoustic axes for piezoelectric media of the symmetry class mm2.

## Introduction

Normals along which a sound wave with polarization vectors perpendicular to the propagation direction could propagate are called specific directions of propagation. They could be divided into longitudinal normals, for which the polarization vector of at least one mode is parallel to the propagation direction and acoustic axes, for which phase velocities of at least two modes are equal. Acoustic axes are also sometimes called degenerate directions or singularities. The study of the number and orientation of specific directions of propagation is important for the proper description of basic acoustic properties of the medium.For this reason they have been the subject of theoretical research for many years [1-27].

Borgnis investigated the number and orientation of longitudinal normals for crystals of trigonal, hexagonal, tetragonal and cubic symmetry [2]. Khatkevich made research of specific directions of propagation for non-piezoelectric media belonging to all classes of symmetry except triclinic one and derived necessary and sufficient conditions which has to be satisfied by a normal to be an acoustic axis. The possible numbers of acoustic axes and longitudinal normals for various types of acoustic symmetry according to Khatkevich [3] are collected in the below given table

| Symmetry of medium | Possible number of longitudinal normals | Possible number Of acoustic axes |
|---|---|---|
| Monoclinic | 17, 15, 13, 11, 9, 7, 5, 3 | 16, 14, 12, 10, 8, 6, 4, 2 |
| Rhombic | 13, 9, 7, 5, 3 | 14, 10, 6, 2 |
| Trigonal | 13, 7 | 10, 4 |
| Tetragonal | 13, 9, 5 | 9, 5, 1 |
| Cubic | 13 | 7 |

Table 1. Given by Khatkevich numbers of longitudinal normals and acoustic axes for non-piezoelectric media of various types of acoustic symmetry

Brugger tabulated polarization vectors and wave speeds for media possessing rhombic, tetragonal, hexagonal and cubic symmetry [4]. The problem of the minimal number of longitudinal normals for non-piezoelectric media of arbitrary symmetry was investigated by Kolodner, Truesdell and Fedorov [5-7]. Sovani et al [8] analyzed the problem of longitudinal normals situated in planes of symmetry of elastic medium or in planes perpendicular to symmetry axes. Sotskaya conducted thorough analysis of the number and orientation of longitudinal normals and acoustic axes for non-piezoelectric media of rhombic symmetry. She claimed that media belonging to this class of acoustic symmetry could possess 2, 6, 10 or 14 distinct acoustic axes [10]. Khatkevich's conditions for the existence of an acoustic axis were reformulated and developed by Alshitz [12-13]. Musgrave  showed that 16 is maximal possible number of acoustic axes for non-piezoelectric media of rhombic symmetry and that rhombic media with 12 distinct acoustic axes could exist [14]. A research of the behaviour of polarization vectors of sound waves propagating in the vicinity of longitudinal normals has been made by Helbig [16]. Bestuzheva and Darinskii have shown that topological indexes of longitudinal normals could provide a basis for examining conditions of their generation and disappearance [17]. The fact that any longitudinal normal is a direction for which the energy flux vector is collinear with the propagation direction has been proved by Ditri. He discussed also the possible application of these results to the determination of the elastic moduli of anisotropic media [18]. Bogdanov investigated longitudinal normals situated inside planes of coordinate system for piezoelectric crystals belonging to the symmetry class *mm2* [19]. Schoenberg and Helbig reconsidered the problem of acoustic axes for media of rhombic symmetry satisfying conditions of so called "mild anisotropy" [20]. Mozhaev et al [22] corrected Khatkevich's results concerning acoustic axes for media of trigonal symmetry. Shuvalov [23] showed that properties of the set of special directions of propagations determine configuration of polarization field of sound waves propagating in this medium. Geymonat and Gilormini [24] provided a new simple proof of Kolodner's statement that every non-piezoelectric medium must have at least 3 longitudinal normals.

Piezoelectricity in principle could increase the possible number of specific directions of propagation. Recently the problem of longitudinal normals for piezoelectric media of the symmetry class mm2 situated inside planes of coordinate system has been analyzed in the interesting paper of Bogdanov [19].

In the present paper we give at the first time examples of rhombic non-piezoelectric media possessing 4 or 8 distinct acoustic axes as well as an example of monoclinic non-piezoelectric medium without acoustic axes. We correct some Bogdanov’s results concerning the longitudinal normals for piezoelectric media belonging to the symmetry class mm2. We discuss also the general method of determination of acoustic axes for piezoelectric media of arbitrary symmetry and apply it to the crystals belonging to the symmetry class mm2.

## Longitudinal normals for piezoelectric media of symmetry class mm2 situated inside symmetry planes

The problem of longitudinal normals for piezoelectric media of symmetry class mm2 was firstly investigated by Bogdanov [19]. According to him longitudinal normals for media of this symmetry contained in the plane (010) are determined by the equation

$$a_1x^2 + a_2x + a_3 = 0, \qquad (1)$$

where

$$
\begin{aligned}
&x = \cos^2 \varphi, \\
&a_1 = \frac{p^2 + \Delta\varepsilon(c_1 + c_2)}{\varepsilon_{33}}, \\
&a_2 = -\left( c_1 + c_2 + \frac{p^2 - p(e_{15} + e_{33}) + \Delta\varepsilon c_2}{\varepsilon_{33}} \right), \\
&a_3 = c_2 - \frac{e_{33}(p - e_{15})}{\varepsilon_{33}},
\end{aligned}
\tag{2}
$$

where φ is the angle between the OX axis and propagation direction and

$$\Delta\varepsilon = (\varepsilon_{33} - \varepsilon_{11}), \tag{3}$$

$$p = 2e_{15} + e_{31} - e_{33}, \tag{4}$$

$$c_1 = C_{11} - C_{13} - 2C_{55}, \tag{5}$$

$$c_2 = C_{22} - C_{13} - 2C_{55}, \tag{6}$$

Components of longitudinal normals situated inside plane (100) could be obtained by appropriate cyclic permutation of indices of materials constants.

We have checked that derived by Bogdanov equation (6) contains small mistake - its correct form is

$$c_2 = C_{33} - C_{13} - 2C_{55}. \tag{7}$$

In the consequence also determined by Bogdanov longitudinal normals situated inside planes (010) and (100) for the following set of material constants (in SI units):

$$
\begin{aligned}
&C_{11} = 15.0\times10^{10},\ C_{22} = 13.4\times10^{10},\ C_{33} = 14.4\times10^{10}, \\
&C_{12} = 2.1\times10^{10},\ C_{13} = 2.6\times10^{10},\ C_{23} = 3.2\times10^{10}, \\
&C_{44} = 4.9\times10^{10},\ C_{55} = 5.2\times10^{10},\ C_{66} = 5.0\times10^{10}, \\
&\mathrm{e}_{15} = 0.82, \mathrm{e}_{24} = 0.73, \mathrm{e}_{31} = 0.26, \mathrm{e}_{32} = 0.17, \mathrm{e}_{33} = 0.47, \\
&\frac{\varepsilon_{11}}{\varepsilon_0} = 6, \frac{\varepsilon_{22}}{\varepsilon_0} = 7, \frac{\varepsilon_{33}}{\varepsilon_0} = 8, \varepsilon_0 = 8.85\times10^{12}.
\end{aligned}
\tag{8}
$$

aren't correct. Bogdanov has found that the plane (010) in the quadrant for which $n_1 > 0$ and $n_2 > 0$ contains two longitudinal normals forming with the axis OX angles 33.6 degrees and 51.6 degrees whereas we found that this plane contains just one longitudinal normal forming with the axis OX the angle 36.8 degrees. For the plane (100) in the quadrant for which $n_3 > 0$ and $n_2 > 0$ we found one longitudinal normal with Cartesian coordinates [0, 0.882, 0.471] - according to Bogdanov this plane doesn't contain any longitudinal normals.

## Longitudinal normals for media of symmetry class mm2 situated outside planes of coordinate system

If **n** is a longitudinal normal situated outside OXY plane and $\Gamma_{ij}$ are elements of the propagation matrix then following equation is satisfied

$$\sum_{j=1}^{3} \Gamma_{ij}\, n_j = \lambda n_i, \tag{9}$$

where $\lambda$ is a real positive number. It could be written in the form

$$[n_1, n_2, n_3] = [\frac{1}{\lambda}\sum_{j=1}^{3}\Gamma_{1j}n_j, \frac{1}{\lambda}\sum_{j=1}^{3}\Gamma_{2j}n_j, \frac{1}{\lambda}\sum_{j=1}^{3}\Gamma_{3j}n_j]. \tag{10}$$

Vectors at both sides of above equations are unit vectors. If $[a_1, a_2, a_3]$ and $[b_1, b_2, b_3]$ are unit vectors with non-zero third component then vectors are equal if and only if following equation connecting ratios of these vectors holds

$$\frac{b_1}{b_3} = \frac{a_1}{a_3}, \frac{b_2}{b_3} = \frac{a_2}{a_3}. \tag{11}$$

Thus (10) is equivalent to

$$\frac{\frac{1}{\lambda}\sum_{j=1}^{3}\Gamma_{1j}n_j,}{\frac{1}{\lambda}\sum_{j=1}^{3}\Gamma_{3j}n_j,} = \frac{n_1}{n_3}, \frac{\frac{1}{\lambda}\sum_{j=1}^{3}\Gamma_{2j}n_j,}{\frac{1}{\lambda}\sum_{j=1}^{3}\Gamma_{3j}n_j,} = \frac{n_2}{n_3}. \tag{12}$$

For elastic media of considered symmetry above set of equations is equivalent to

$$\begin{aligned} & A_4 n_1^4 + A_2 n_1^2 + A_0 = 0, \\ & B_4 n_1^4 + B_2 n_1^2 + B_0 = 0. \end{aligned} \tag{13}$$

where

$$\begin{aligned}
& A_4 = -2C_{55}\varepsilon_{11} - e_{31}^2 + C_{11}\varepsilon_{11} - 2e_{15}e_{31} - C_{13}\varepsilon_{11}, \\
& A_2 = (-2C_{55}\varepsilon_{22} - C_{13}\varepsilon_{22} - 2C_{44}\varepsilon_{11} - 2e_{24}e_{31} - 2e_{31}e_{32} - \\
& -C_{23}\varepsilon_{11} + 2C_{66}\varepsilon_{11} + C_{11}\varepsilon_{22} + C_{12}\varepsilon_{11} - 2e_{15}e_{32})n_2^2 + \\
& +(C_{11}\varepsilon_{33} - C_{33}\varepsilon_{11} + 2e_{15}e_{31} - 2e_{15}e_{33} - C_{13}\varepsilon_{33} - 2e_{31}e_{33} \\
& -2C_{55}\varepsilon_{33} + 4e_{15}^2 + 2C_{55}\varepsilon_{11} + C_{13}\varepsilon_{11})n_3^2, \\
& A_0 = (-2C_{44}\varepsilon_{22} - C_{23}\varepsilon_{22} - 2e_{24}e_{32} + 2C_{66}\varepsilon_{22} - e_{32}^2 + C_{12}\varepsilon_{22})n_2^4 + \\
& (-2C_{44}\varepsilon_{33} + 2C_{66}\varepsilon_{33} + 4e_{24}e_{15} - C_{23}\varepsilon_{33} + C_{12}\varepsilon_{33} - 2e_{24}e_{33} - \\
& -2e_{32}e_{33} + 2C_{55}\varepsilon_{22} - C_{33}\varepsilon_{22} + 2e_{15}e_{32} + C_{13}\varepsilon_{22})n_2^2 n_3^2 + \\
& (2e_{15}e_{33} + 2C_{55}\varepsilon_{33} - C_{33}\varepsilon_{33} - e_{33}^2 + C_{13}\varepsilon_{33})n_3^4,
\end{aligned} \tag{14}$$

and

$$\begin{aligned}
& B_4 = -e_{32}^2 - 2e_{24}e_{32} + C_{22}\varepsilon_{22} - C_{23}\varepsilon_{22} - 2C_{44}\varepsilon_{22}, \\
& B_2 = (-2e_{24}e_{31} - C_{13}\varepsilon_{22} - C_{23}\varepsilon_{11} + C_{12}\varepsilon_{22} + C_{22}\varepsilon_{11} - \\
& -2C_{55}\varepsilon_{22} - 2e_{31}e_{32} + 2C_{66}\varepsilon_{22} - 2C_{44}\varepsilon_{11} - 2e_{15}e_{32})n_1^2 + \\
& (-C_{23}\varepsilon_{33} + C_{22}\varepsilon_{33} - 2e_{24}e_{33} + C_{23}\varepsilon_{22} - C_{33}\varepsilon_{22} + 4e_{24}^2 - \\
& -2e_{32}e_{33} + 2e_{24}e_{32} + 2C_{44}\varepsilon_{22} - 2C_{44}\varepsilon_{33})n_3^2, \\
& B_0 = (-2e_{15}e_{31} + 2C_{66}\varepsilon_{11} - 2C_{55}\varepsilon_{11} - e_{31}^2 - C_{13}\varepsilon_{11} + \\
& +C_{12}\varepsilon_{11})n_1^4 + (C_{12}\varepsilon_{33} + 2C_{44}\varepsilon_{11} - 2e_{31}e_{33} + C_{23}\varepsilon_{11} - \\
& -2C_{55}\varepsilon_{33} - C_{33}\varepsilon_{11} + 4e_{15}e_{24} + 2C_{66}\varepsilon_{33} - C_{13}\varepsilon_{33} - 2e_{15}e_{33} + \\
& +2e_{24}e_{31})n_3^2 n_1^2 + (-e_{33}^2 + C_{23}\varepsilon_{33} - C_{33}\varepsilon_{33} + 2C_{44}\varepsilon_{33} + 2e_{24}e_{33})n_3^4.
\end{aligned} \tag{15}$$

The set of equations (10) is satisfied only for these normals **n** for which following equation holds

$$A_0^2 B_4^2 - 2A_0 A_4 B_0 B_4 + A_4^2 B_0^2 - A_0 A_2 B_2 B_4 - A_2 A_4 B_0 B_2 + A_0 A_4 B_2^2 + A_2^2 B_0 B_4 = 0. \tag{16}$$

Using above equations one can determine all free longitudinal normals for considered medium. As an example longitudinal normals for $LiGaO_2$ will be determined. This material is characterized by the following set of materials constants [27]

$$
\begin{aligned}
&C_{11} = 140, C_{22} = 120, C_{33} = 140, C_{44} = 57.1, C_{44} = 47.4,\\
&C_{66} = 69.0, C_{12} = 14, C_{13} = 28, C_{23} = 31,\\
&\frac{\varepsilon_{11}}{\varepsilon_0} = 7.0, \frac{\varepsilon_{22}}{\varepsilon_0} = 6.5, \frac{\varepsilon_{33}}{\varepsilon_0} = 8.3,\\
&e_{15} = -0.32, e_{24} = -0.34, e_{31} = -0.17, e_{32} = -0.31, e_{33} = 0.96, \rho = 4187.
\end{aligned} \tag{17}
$$

By solving above set of equations one obtains that longitudinal normals for considered compound situated outside planes of the coordinate system have following components: [0.458, 0.610, 0.647], [-0.458, 0.610, 0.647], [0.458, -0.610, 0.647], [-0.458, -0.610, 0.647].

## Example of non-piezoelectric medium of monoclinic symmetry without acoustic axes

Alshitz and Lothe have shown that media of rhombic symmetry for which elastic constants satisfy following conditions

$$C_{33} >> C_{55} > C_{44} > C_{11} > C_{66} > C_{22} > 0 \tag{18}$$

haven't any acoustic axes. Boulanger and Hayes [21] gave the example of rhombic medium for which the diagonal elements of the matrix of elastic constants obey inequalities (18) and non-diagonal elements of this matrix are small but non-zero. Following their approach we checked that the monoclinic medium characterized by the following set of elastic constants

$$
\begin{aligned}
&C_{11} = 86.66, C_{22} = 75.45, C_{33} = 4000, C_{44} = 90.45, C_{55} = 95.23\\
&C_{66} = 81.12, C_{12} = -25, C_{13} = -10, C_{23} = -24, C_{16} = 0.9, C_{26} = 0.69,\\
&C_{36} = -1.69, C_{45} = 2.0
\end{aligned} \tag{19}
$$

doesn't possess any acoustic axes.

## Acoustic axes for non-piezoelectric media of rhombic symmetry

As we have mentioned according to Khatkevich and Sotskaya [10, 11] the possible number of acoustic axes for media of rhombic symmetry is 14, 10, 6 or 2. Alshitz has proven that one can find an example of the medium of rhombic symmetry which doesn't possess any acoustic axes [13]. Examples of rhombic media with 12 and 16 distinct acoustic axes have been reported in articles [14, 21]. Below we give examples of rhombic media with 4 or 8 distinct axes.

Let us consider the medium characterized by the following set of elastic constants:

$$\begin{aligned} &C_{11}=86.66, C_{22}=75.45, C_{33}=4000.0, C_{44}=90.45,\\ &C_{55}=95.23, C_{66}=81.12, C_{12}=-25.0, C_{13}=-10.0, C_{23}=-24.0. \end{aligned} \quad (20)$$

This medium possesses only 4 acoustic axes. Their components are [0.2152, 0.9765, 0], [0.2152, -0.9765, 0], [-0.9797, 0.2006, 0] and [-0.9797, -0.2006, 0].

Now we will present the example of the set of elastic constants for which exactly 8 acoustic axes exist

$$\begin{aligned} &C_{11}=107, C_{22}=105, C_{33}=500, C_{44}=108,\\ &C_{55}=109.1, C_{66}=106, C_{12}=6, C_{13}=7, C_{23}=5. \end{aligned} \quad (21)$$

Above medium has 4 acoustic axes situated inside OXY plane and 4 acoustic axes situated inside OXZ plane. Their components are:

[0.02187, 0.99976, 0], [-0.99974, -0.02279, 0], [-0.02187, 0.99976, 0], [0.99974, -0.02279, 0], [-0.1786, 0, -0.9839], [0.1786, 0, -0.9839], [-0.99988, 0, -0.01539] and [0.99988, 0, -0.01539]. The OYZ plane doesn't contain any acoustic axes.

The medium also doesn't possess any acoustic axes situated outside planes of the coordinate system.

In the below table we give corrected version of the table given by Khatkevich in the paper [3] and collect possible number of longitudinal normals and acoustic axes for non-piezoelectric media

| Symmetry of medium | Possible number of longitudinal normals | Possible number of acoustic axes |
|---|---|---|
| Monoclinic | 13, 11, 9, 7, 5, 3 | 16, 14, 12, 10, 8, 6, 4, 2, 0 |
| Rhombic | 13, 9, 7, 5, 3 | 16, 14, 12, 10, 8, 6, 4, 2, 0 |
| Trigonal | 13, 7 | 16, 10, 4 |
| Tetragonal | 13, 9, 5 | 9, 5, 1 |
| Cubic | 13 | 7 |

Table 2. Number of longitudinal normals and acoustic axes for non-piezoelectric media for various types of acoustic symmetry

There are following differences between above table and the table presented by Khatkevich:

1. For media of trigonal symmetry one can have 16 acoustic axes as the implication of considerations presented in the paper of Mozhaev et al. [22].
2. Maximal number of longitudinal normals for monoclinic media is 13, not 17 [26].
3. Alshitz and Lothe have shown that media of rhombic symmetry for which elastic constants satisfy following conditions

$$C_{33} >> C_{55} > C_{44} > C_{11} > C_{66} > C_{22} > 0 \qquad (22)$$

haven't any acoustic axes - thus the smallest possible number of acoustic axes for media of rhombic symmetry is 0 [13] .

4. In articles [14, 21] examples of rhombic media with 12 or 16 distinct acoustic axes were given.
5. In this chapter examples of rhombic media with 4 and 8 distinct acoustic axes have been presented.
6. We gave above an example of monoclinic medium without acoustic axes.

## Acoustic axes for piezoelectric media

The effect of piezoelectricity on the propagation of acoustic waves in elastic media has been the subject of research of many authors. In particular Lyubimov [9] investigated the influence of piezoelectricity on the phase velocity of sound wave and gave the necessary and sufficient condition which has to be satisfied by the normal **n** if it is
an acoustic axis. According to him the direction **n** is an acoustic axis if and only if it satisfies given below equation

$$4F_1^3(n) - F_2^2(n) = 0, \qquad (23)$$

where $F_1(n)$ and $F_2(n)$ are complicated polynomial functions depending on components of the propagation direction ***n*** and elements of tensors of elastic, piezoelectric and dielectric constants - we don' t cite here their explicit form since they are very complicated. For given set of materials constants the function $F_1(n)$ is a homogeneous polynomial function in components of the propagation direction of eighth degree whereas the degree of the function $F_2(n)$ is 12 - thus the degree of the equation is 24.

Although Lyubimov hasn't discussed in his article directly the problem of determination of components of degenerate directions nor of the maximal possible number of these directions, derived by him equation (23) in principle could be applied to solve these problems.

Every elastic medium with symmetry other than isotropic or hexagonal generally has finite number of acoustic axes. Since according to Lyubimov the direction ***n*** is an acoustic axis if and only if the equation (23) for this normal is satisfied - so for media possessing finite number of acoustic axes the number of solutions of this equation would also be finite. Thus also the number of roots of the function standing at the left side of this equation is also finite. In the reference [26] it has been demonstrated that if for a continuous function $F(x,y)$ defined on $R^2$ in the set of values of this function one can find at least one pair of numbers with different signs then this function has infinite many roots as the consequence of Roll's theorem. In the same way one can prove that the fact that the number of roots of the function

$$G(n) \equiv 4F_1^3(n) - F_2^2(n) \qquad (24)$$

is finite implies that all values of this function are nonpositive or all are nonnegative, thus all roots of the equation (23) have multiplicity higher than one. On the other hand the degree of the considered equation is high - for this reason we think that solving this equation would involve serious numerical difficulties, especially for media of triclinic or monoclinic symmetry. In any way in the known to us literature it hasn't been demonstrated in practice that this method could be practically applied for the determination of degenerate directions for these media and Lyubimov's article wasn't aimed at tackling this problem. For this reason we propose here different approach based on methods used previously for the determination of components of longitudinal normals for non-piezoelectric media [26].

The direction **n** is an acoustic axis if and only if the elements of the propagation for this direction satisfy one from the below given conditions [3]:

**Condition A**

The following set of equations is satisfied

$$\begin{cases}(\Gamma_{11}-\Gamma_{22})\Gamma_{23}\Gamma_{13}+\Gamma_{12}(\Gamma_{23}^2-\Gamma_{13}^2)=0\ , \\ (\Gamma_{11}-\Gamma_{33})\Gamma_{23}\Gamma_{12}+\Gamma_{13}(\Gamma_{23}^2-\Gamma_{12}^2)=0.\end{cases} \tag{25}$$

and all non-diagonal elements of the propagation matrix are non-zero.

**Condition B**

At least one from the following sets of equations is satisfied

$$\begin{cases}\Gamma_{23}=0, \\ \Gamma_{13}=0, \\ (\Gamma_{11}-\Gamma_{33})(\Gamma_{22}-\Gamma_{33})-\Gamma_{12}^2=0\ ,\end{cases} \tag{26}$$

$$\begin{cases}\Gamma_{12}=0, \\ \Gamma_{23}=0, \\ (\Gamma_{33}-\Gamma_{22})(\Gamma_{11}-\Gamma_{22})-\Gamma_{13}^2=0,\end{cases} \tag{27}$$

$$\begin{cases}\Gamma_{13}=0, \\ \Gamma_{12}=0, \\ (\Gamma_{33}-\Gamma_{11})(\Gamma_{22}-\Gamma_{11})-\Gamma_{23}^2=0.\end{cases} \tag{28}$$

For piezoelectric media components of the propagation matrix have the following form:

$$\Gamma_{il}=\rho^{-1}\sum_{j.m=1}^{3}\left(C_{ijlm}^{E}+\frac{\sum_{r,s=1}^{3}e_{ijr}n_r e_{lms}n_s}{\sum_{p,q=1}^{3}\varepsilon_{pq}n_p n_q}\right)n_j n_m \tag{29}$$

where ρ is the density of medium, $C_{ijlm}^{E}$ is the tensor of elastic constants, $e_{ijr}$ is the tensor of piezoelectric constants and $\varepsilon_{pq}$ is the tensor of dielectric constants.

Let us assume that $P_r(\Gamma_{ij})$ is a homogeneous polynomial function of the elements of the propagation matrix for piezoelectric medium and that the degree of this polynomial function is *r*. In this case for the given set of material constants the equation

$$P_r(\Gamma_{ij})=0 \tag{30}$$

could be transformed to the homogeneous polynomial equation in variables $n_1, n_2, n_3$ of the degree *4r* by multiplying both sides of (35) by $(\sum_{p,q=1}^{3} \varepsilon_{pq} n_p n_q)^r$. This equation could be easily solved by transformation to variables $x \equiv \frac{n_1}{n_3}, y \equiv \frac{n_2}{n_3}$ and solving obtained set of polynomial equations with the help of the method applied previously by us to the determination of components of longitudinal normals for non-piezoelectric media [26].

Since left hands of all sets (25)-(28) are homogeneous polynomial functions of components of the propagation matrix for piezoelectric media, one can apply this method for the determination of components of acoustic axes. The method could be used for media of any symmetry - in the next subsections we make use of it for the analysis of acoustic axes for media of the symmetry class mm2.

To demonstrate that described above method could be practically used for the computation of acoustic axes we will determine acoustic axes for $Li_2SO_4{*}H_2O$

$$\begin{aligned}
&C_{11} = 54.9, C_{22} = 70.5, C_{33} = 61.8, C_{44} = 14.0, C_{55} = 24.2,\\
&C_{66} = 27.0, C_{12} = 26.3, C_{13} = 11.4, C_{23} = 17.1, C_{15} = 6.5, C_{25} = 15.7, C_{35} = -5.2, C_{46} = -26.5\\
&\frac{\varepsilon_{11}}{\varepsilon_0} = 5.6, \frac{\varepsilon_{22}}{\varepsilon_0} = 10.3, \frac{\varepsilon_{33}}{\varepsilon_0} = 6.5,\\
&e_{14} = -0.052, e_{16} = -0.083, e_{21} = 0.238, e_{22} = 0.81, e_{23} = 0.171, e_{25} = -0.015, e_{34} = -0.015, , e_{36} = 0.031..
\end{aligned} \qquad (31)$$

Following the procedure described above we found that following normals are acoustic axes: [0.191, -0.276, 0.942], [0.191, 0.276, 0.942], [0.744, 0.617, 0.258], [0.744, -0.617, 0.258] and [-0.127, 0, 0.992].

## Acoustic axes for rhombic media of symmetry class mm2 situated inside planes of coordinate system

One can easily check that for media of symmetry class mm2 the condition A is never satisfied.

In the OXZ plane $\Gamma_{12}$ and $\Gamma_{23}$ vanish and the direction **n** is an acoustic axis if and only if the following set of equations is satisfied for this direction

$$\sum_{i=0}^{4} A_i^{XZ} n_1^{2(4-i)} n_2^{2i} = 0 \qquad (32)$$

and $A_0^{XZ}, A_1^{XZ}, A_2^{XZ}, A_3^{XZ}, A_4^{XZ}$ are functions of material constants characterizing the medium - we don't write them here explicitly due to their complicated form.

Analogous conditions satisfy acoustic axes situated inside OXY plane and OYZ plane.

**Acoustic axes for rhombic media of symmetry class mm2 situated outside planes of coordinate system**

Outside planes of coordinate system the set (30) for media of symmetry class mm2 could be written in the form

$$\sum_{i,j=1}^{9} A_{ij} n_1^i n_2^j n_3^{9-j} = 0, \tag{33}$$

$$\sum_{i,j=1}^{9} B_{ij} n_1^i n_2^j n_3^{9-j} = 0, \tag{34}$$

where $A_{ij}$ and $B_{ij}$ are complicated functions depending only on the material constants characterizing the medium.

One can easily check that the normal **n** for piezoelectric media belonging to the symmetry class mm2 lying out of symmetry planes is an acoustic axis if and only if the set (33)-(34) for this medium is satified and simultaneously one can't find for this normal a pair of non-diagonal elements of the propagation matrix equal zero. Thus the normal **n** situated outside planes of coordinate system is an acoustic axis if and only if the set (33)-(34) is fulfilled and none from the below sets is satified

$$\begin{cases} (C_{44}\varepsilon_{33}+C_{23}\varepsilon_{33}+2e_{24}e_{33})n_3^2+(C_{44}\varepsilon_{22}+C_{23}\varepsilon_{22}+2e_{24}e_{32})n_2^2+(C_{44}\varepsilon_{11}+C_{23}\varepsilon_{11}+2e_{24}e_{31})n_1^2=0, \\ (C_{55}\varepsilon_{33}+C_{13}\varepsilon_{33}+2e_{15}e_{33})n_3^2+(C_{55}\varepsilon_{11}+C_{13}\varepsilon_{11}+2e_{15}e_{32})n_2^2+(C_{55}\varepsilon_{11}+C_{13}\varepsilon_{22}+2e_{15}e_{31})n_1^2=0, \end{cases} \tag{35}$$

$$\begin{cases} (C_{44}\varepsilon_{33}+C_{23}\varepsilon_{33}+2e_{24}e_{33})n_3^2+(C_{44}\varepsilon_{22}+C_{23}\varepsilon_{22}+2e_{24}e_{32})n_2^2+(C_{44}\varepsilon_{11}+C_{23}\varepsilon_{11}+2e_{24}e_{31})n_1^2=0, \\ (C_{66}\varepsilon_{33}+C_{12}\varepsilon_{33}+4e_{15}e_{24})n_3^2+(C_{66}\varepsilon_{22}+C_{12}\varepsilon_{22})n_2^2+(C_{66}\varepsilon_{11}+C_{12}\varepsilon_{11})n_1^2=0, \end{cases} \tag{36}$$

$$\begin{cases} (C_{66}\varepsilon_{33}+C_{12}\varepsilon_{33}+4e_{15}e_{24})n_3^2+(C_{66}\varepsilon_{22}+C_{12}\varepsilon_{22})n_2^2+(C_{66}\varepsilon_{11}+C_{12}\varepsilon_{11})n_1^2=0, \\ (C_{55}\varepsilon_{33}+C_{13}\varepsilon_{33}+2e_{15}e_{33})n_3^2+(C_{55}\varepsilon_{11}+C_{13}\varepsilon_{11}+2e_{15}e_{32})n_2^2+(C_{55}\varepsilon_{11}+C_{13}\varepsilon_{22}+2e_{15}e_{31})n_1^2=0, \end{cases} \tag{37}$$

Left sides of above sets of equations are appropriate non-diagonal elements of the propagation matrix for piezoelectric media of the symmetry class mm2 multiplied by $\sum_{p,q=1}^{3} \varepsilon_{pq} n_p n_q$.

## Example

As an example we will determine acoustic axes for $LiGaO_2$. Using above equations I checked that considered medium hasn't any acoustic axes situated outside planes of coordinate system. Inside OYZ it has following acoustic axes: [0, 0.577, 0.817] and [0, 0.577, 0.817].